\documentclass[preprint,sd,amsmath,amssymb,graphicx,superscriptaddress]{revtex4-1}

\usepackage{graphicx}
\usepackage{dcolumn}
\usepackage{bm}
\usepackage{SIunits}

\draft 

\begin{document}


\title{A low disorder Metal-Oxide-Silicon double quantum dot} 



\author{J.-S. Kim}
\email[]{jk3@princeton.edu}
\affiliation{Department of Electrical Engineering, Princeton University, Princeton, NJ 08544, USA}
\author{T. M. Hazard}
\affiliation{Department of Electrical Engineering, Princeton University, Princeton, NJ 08544, USA}
\author{A. A. Houck}
\affiliation{Department of Electrical Engineering, Princeton University, Princeton, NJ 08544, USA}
\author{S. A. Lyon}
\email[]{lyon@princeton.edu}
\affiliation{Department of Electrical Engineering, Princeton University, Princeton, NJ 08544, USA}


\date{\today}

\begin{abstract}
 
One of the biggest challenges impeding the progress of metal-oxide-silicon (MOS) quantum dot devices is the presence of disorder at the Si/SiO$_2$ interface which interferes with controllably confining single and few electrons. In this work we have engineered a low-disorder MOS quantum double-dot device with critical electron densities, i.e. the lowest electron density required to support a conducting pathway, approaching critical electron densities reported in high quality Si/SiGe devices and commensurate with the lowest critical densities reported in any MOS device. Utilizing a nearby charge sensor, we show that the device can be tuned to the single-electron regime where charging energies of $\approx$8 meV are measured in both dots, consistent with the lithographic size of the dot. Probing a wide voltage range with our quantum dots and charge sensor, we detect three distinct electron traps, corresponding to a defect density consistent with the ensemble measured critical density. Low frequency charge noise measurements at 300 mK indicate a 1/$f$ noise spectrum of 3.4 $\micro$eV/Hz$^{1/2}$ at 1 Hz and magnetospectroscopy measurements yield a valley splitting of 110$\pm$26 $\micro$eV. This work demonstrates that reproducible MOS spin qubits are feasible and represents a platform for scaling to larger qubit systems in MOS.

\end{abstract}


\maketitle 


\section{Introduction}

Electron spins in silicon devices are promising qubits for a quantum processor, defining a natural two-level system and demonstrating long spin coherence times \cite{tyryshkin, 28ms_T2, 30s_T2}. Recently, 2-qubit quantum controlled-not (CNOT) operations have been demonstrated in both Si/SiGe \cite{zajac_cnot} and Si Metal-Oxide-Semiconductor (MOS) quantum dot systems \cite{veldhorst_cnot}, establishing a crucial building block for a universal quantum computer \cite{divincenzo_criterion}. While the MOS system allows for electron-donor interactions (enabling an avenue for quantum memories in donor states)\cite{morello2012,harveycollard} and demonstrates larger valley splittings (critical for high fidelity spin-selective operations) \cite{yang_vs,gamble_vs,xiao_vs,zajac_dqd, borselli_vs}, it suffers from high disorder compared to its Si/SiGe counterpart \cite{zwanenburg}. Indeed, because of disorder, the scaling-up of MOS quantum dots to multi-qubit systems has lagged behind Si/SiGe systems, where an array of nine uniform quantum dots have been achieved \cite{zajac_9qd} as well as the coupling of a single spin to a superconducting resonator \cite{mi_spin_photon,samkharadze}. Furthermore, in MOS, disorder may be unintentionally introduced to the Si/SiO$_2$ interface during high-energy processes like electron-beam lithography \cite{kim_sio2, nordberg}, an essential fabrication process for quantum dot devices produced by research labs. In this work we address the issue of disorder in MOS quantum dots by engineering and characterizing a low-disorder MOS quantum dot device where the disorder parameters critical for quantum dot devices, i.e. density of shallow traps and critical density \cite{kim_sio2, jock}, approach the low-disorder levels demonstrated in Si/SiGe systems \cite{zajac_dqd, mi_mu}.

The metric typically cited to characterize disorder in quantum dot devices is the low-temperature electron mobility ($\mu$), where Si/SiGe devices routinely report mobilities from 10$^5$ to 10$^6$ cm$^2$/Vs, up to two orders of magnitude higher than the best MOS devices \cite{kim_sio2, kravchenko, nordberg, tracy, mi_mu, borselli_mu, zajac_dqd, xiao_mu}.  However, the peak mobility is an insufficient metric for quantum dot quality because the peak mobility occurs at high electron densities ($\sim$10$^{12}$ cm$^{-2}$) where enough electrons are present in the system to screen defects and disorder \cite{ando}. Quantum dot devices operate in the single or few electron regime ($\sim$10$^{10}$ cm$^{-2}$) where the peak mobility value is less applicable. Instead, the more appropriate metrics of disorder in quantum dot devices are 1) the critical density ($n_0$), i.e. the lowest electron density required to support a conducting pathway, and 2) the density of shallow electron traps, i.e. electrically active electron traps within a few meV of the conduction band edge and present at cryogenic temperatures \cite{kim_sio2, jock}.

\section{Results}

We have fabricated and characterized a low disorder MOS quantum double-dot device, leveraging a previously published process yielding very low critical and shallow trap densities (8.3 - 9.5 $\times$ 10$^{10}$ cm$^{-2}$), and simultaneously very high mobilities (1.4 - 2.3 $\times 10^{4}$ cm$^2$/Vs), despite exposure to high-energy processes like electron-beam lithography \cite{kim_sio2}. We note that the critical densities of this starting gate stack are within a factor of 2-3 of critical densities reported in Si/SiGe devices (4.6 $\times$ 10$^{10}$ cm$^{-2}$) \cite{zajac_dqd} and are on par with the lowest critical densities reported in MOS \cite{kravchenko} (which used thick-oxide devices). We adopt a reconfigurable device architecture pioneered in Si/SiGe devices \cite{zajac_dqd}, with three overlapping layers of gates defining two parallel conduction channels. Notably, our device's first layer of gates are fabricated from degenerately doped poly-silicon instead of the typical aluminum. Low frequency bias spectroscopy measurements through each individual dot in the upper channel show regular Coulomb blockade diamonds, demonstrating low levels of disorder and the charging of a single lithographically defined dot. We define a charge sensor dot in the center of the lower conduction channel and detect the depletion of each upper channel quantum dot to the single-electron regime and additionally show the controllable formation of a quantum double-dot. Our charge sensor also detects the charging of individual poly-silicon grains (``phantom dots") within the poly-silicon depletion gates which do not affect the charge state of our quantum dots but add a prominent background signal. Charge-noise spectroscopy measurements yield a 1/$f$ spectrum in power spectral density with a value of 3.4 $\micro$eV/Hz$^{1/2}$ at 1 Hz, consistent with other reported values of charge noise in MOS and Si/SiGe devices \cite{jiang_noise}. Monitoring the N=0$\rightarrow$1 and N=1$\rightarrow$2 transition as a function of a perpendicular magnetic field, we measure a valley splitting of 110$\pm$26 $\micro$eV.

\begin{figure}
\includegraphics[width=0.7\linewidth]{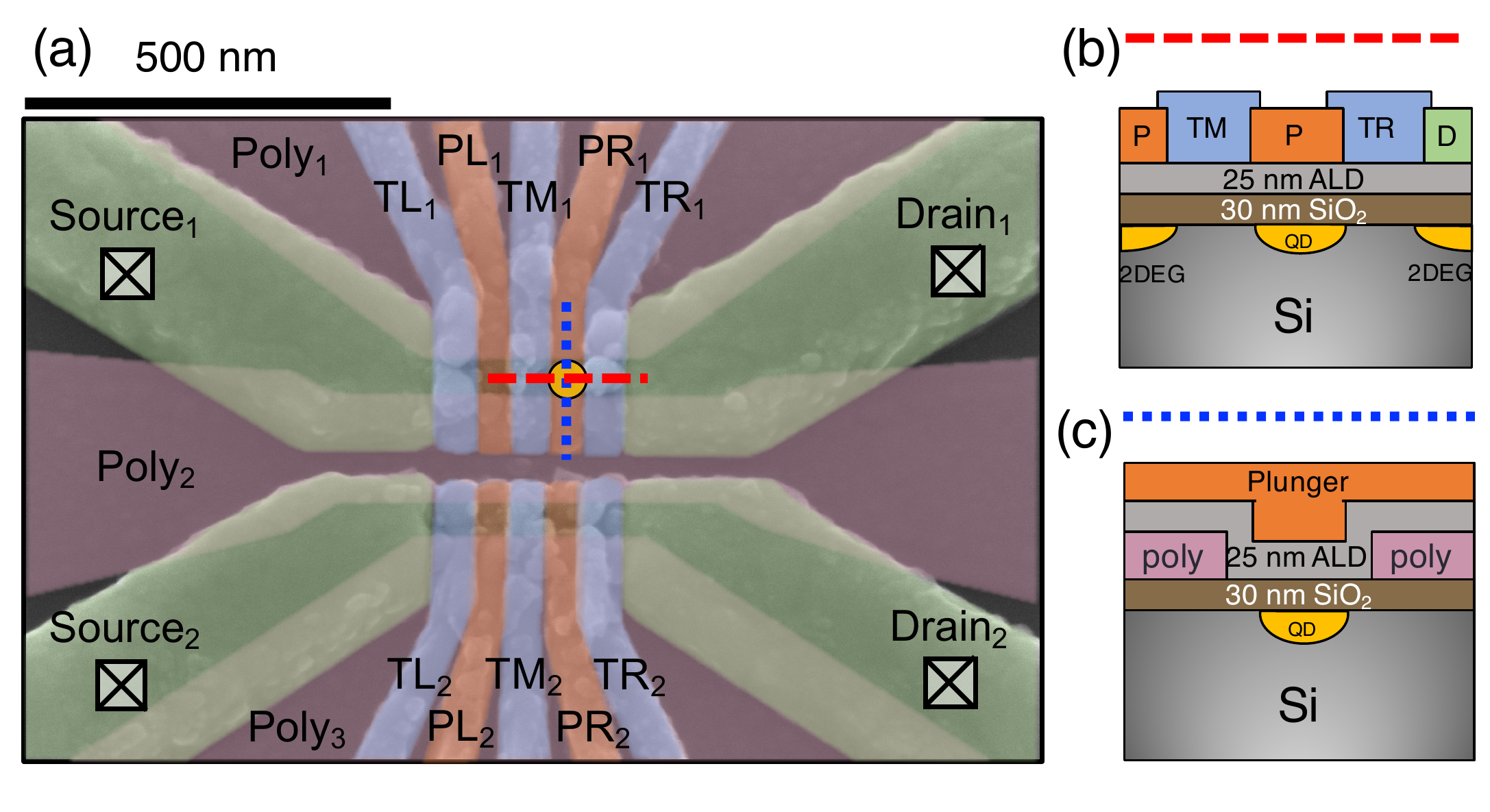}
\caption{\label{device}(a) False color SEM micrograph of a device identical to the one measured. Schematic of (b) horizontal and (c) vertical cross-sections of the device as indicated by the dotted lines in (a).}
\end{figure}

The device fabrication process was engineered to minimize shallow traps by minimizing high-energy processes and eliminating vectors of mobile ionic contamination. Starting from a previously published MOSFET gate stack (1,000-3,000 $\Omega$-cm p-type (100) silicon, 30 nm dry, chlorinated thermal oxide, 200 nm n+ poly-Si gate), \cite{kim_sio2} we adopt an overlapping gate architecture \cite{zajac_dqd} capable of defining up to four individual quantum dots with independent control of each dot's electron occupation, tunnel barriers and source/drain reservoirs (Figure \ref{device}). The first layer of gates is achieved by thinning the poly-Si gate down to 40 nm with a low power reactive ion etch step (50 W, SF$_6$ and C$_4$F$_8$) and etching two parallel 80 nm wide channels, defined by electron-beam (e-beam) lithography (Elionix F-125), down to the SiO$_2$. These channels define three poly-Si depletion gates (labelled Poly$_{1,2,3}$ in Fig. \ref{device}(a)) which serve as a screening layer for the subsequent Al accumulation gates and also prevent the diffusion of mobile ionic contaminants into the underlying gate oxide during the fabrication process. Immediately after stripping the e-beam resist, 200 cycles of atomic layer deposition aluminum oxide (ALD) were grown on the sample to insulate the poly-Si gates from the subsequent Al gates and to protect the exposed oxide from contamination \cite{ald_sio2}. Two subsequent layers of 50 nm wide aluminum gates were deposited via lift-off, defining source/drain reservoirs (labelled Source$_{1/2}$, Drain$_{1/2}$), plunger gates (labelled PL$_{1/2}$, PR$_{1/2}$), and tunnel barrier gates (labelled TL$_{1/2}$, TM$_{1/2}$, TR$_{1/2}$). These gates were oxidized between depositions in an O$_2$ plasma for electrical insulation \cite{angus}. Finally the device was furnace annealed at 400$\degree$ C in forming gas (95\% N$_2$, 5\% H$_2$) for 30 minutes to lower the interface state density \cite{kim_sio2,nicbrews}.

One notable change in the materials in this gate stack and the low-disorder starting gate stack is the addition of ALD aluminum oxide on top of the thermal gate SiO$_2$. The introduction of new materials to the gate stack can alter the quality of the Si/SiO$_2$ interface as the SiO$_2$/ALD interface can contain substantial amounts of fixed negative charge \cite{ald_sio2}. However, Van der Pauw devices fabricated with this new Si/SiO$_2$/ALD/Al gate stack achieved low electron densities ($\sim 1.0\times 10^{11}$ cm$^{-2}$ at 4.2 K) and high mobilities (1.2-1.4 $\times$ 10$^{4}$ cm$^2$/Vs) so we are confident that the effect of the ALD on the Si/SiO$_2$ interface disorder is minimal.

For this work, we bias the device to form two quantum dots in the upper conduction channel and define a charge sensor dot in the center of the lower conduction channel. All measurements were taken at a base temperature of 305 mK in a Janis $^3$He cryostat with a corresponding electron temperature of $\sim$305 mK estimated by the line width of the Coulomb blockade peaks.

\begin{figure}
\includegraphics[width=0.7\linewidth]{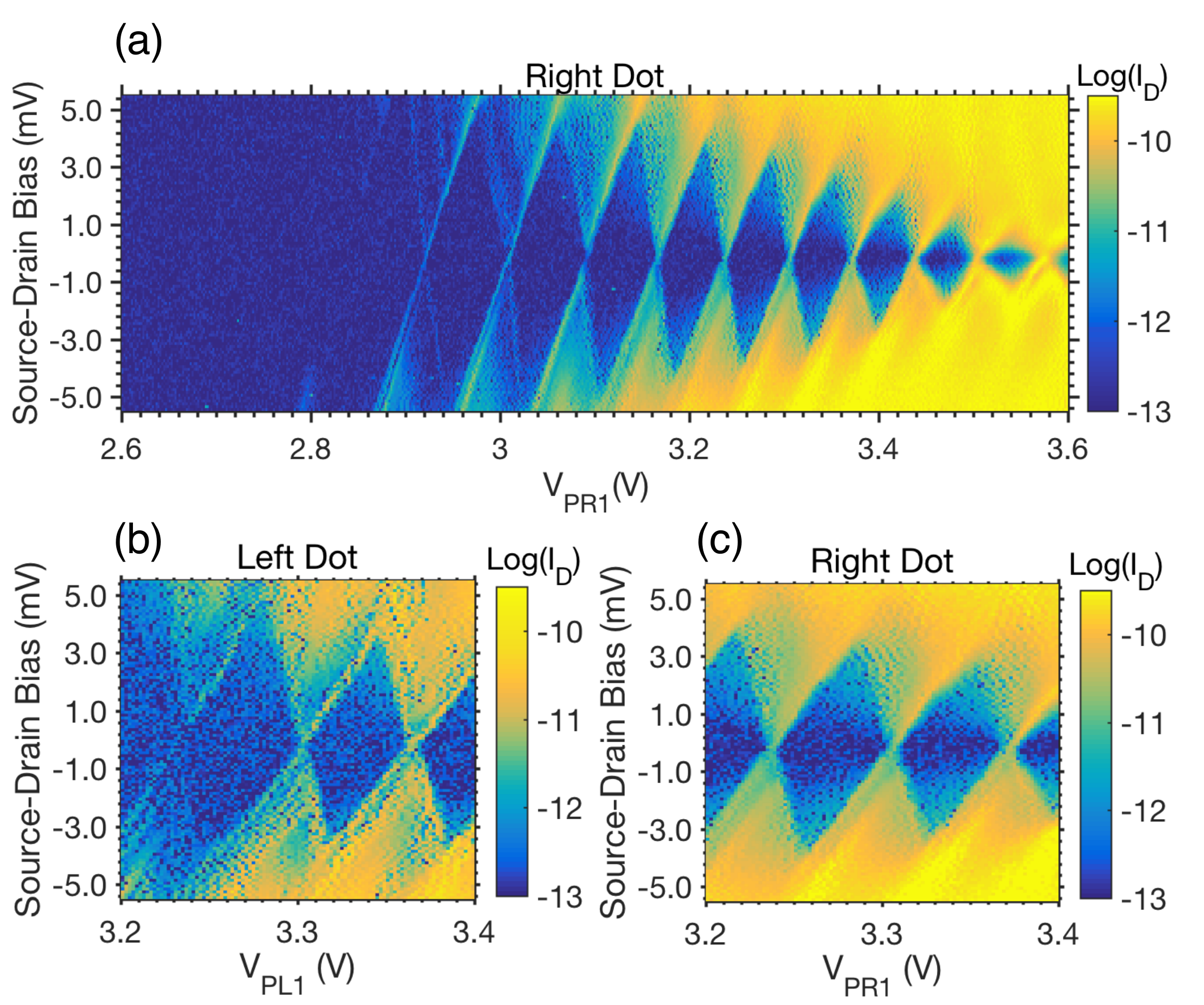}
\caption{\label{diamond} (a) Low frequency conductance measurements of the upper right quantum dot over a wide parameter range. Regular Coulomb blockade diamonds monotonically increase in size and open up at low plunger voltages, indicating the dot has reached the few-electron regime. Comparison of the left (b) and right (c) Coulomb blockade diamonds demonstrate good uniformity and charging energies consistent with the lithographic size of the dots.}
\end{figure}

We first characterize each quantum dot individually with low-frequency bias spectroscopy measurements to extract the charging energies and lever arms of each dot. Measurements were taken using an Ithaco 1211 current preamplifier in conjunction with an SR830 lock-in amplifier at a frequency of 229 Hz and an AC excitation amplitude of 50 $\micro$V. We observe regular Coulomb blockade diamonds over a wide electron occupation range which monotonically decrease in size with increasing electron occupation, reflecting the increase in quantum dot size with increasing voltage applied to the plunger gate (Fig. \ref{diamond} (a)). At lower plunger voltages, the Coulomb diamonds appear to cease, indicating the dot has reached the few-electron regime. In this regime, the charging energies ($E_C$) for both dots are measured to be $\approx$5 meV, and we extract lever arms of 0.059 meV/mV for the right dot and 0.067 meV/mV for the left dot, respectively (Fig. \ref{diamond}(b), (c)). 

As the plunger voltage is made more and more negative to depopulate the quantum dot, the tunnel barriers simultaneously become more opaque due to the cross capacitance of the plunger gate to the tunnel barriers. Thus, as the quantum dot is depopulated, the tunnel rate through the dot decreases, resulting in a conductance signal which falls below the noise floor of the measurement system. In order to interrogate the single-electron regime, we define a charge sensor dot in the center of the lower conduction channel beneath TM$_2$. When the charge sensor dot is biased to the edge of a Coulomb blockade peak, small changes to the local electrostatic environment (e.g. the addition of single electrons in the upper quantum dots) result in a measurable change in current through the charge sensor dot. With the charge sensor, we may explicitly show that the upper quantum dots have reached the single-electron regime.

\begin{figure}
\includegraphics[width=0.7\linewidth]{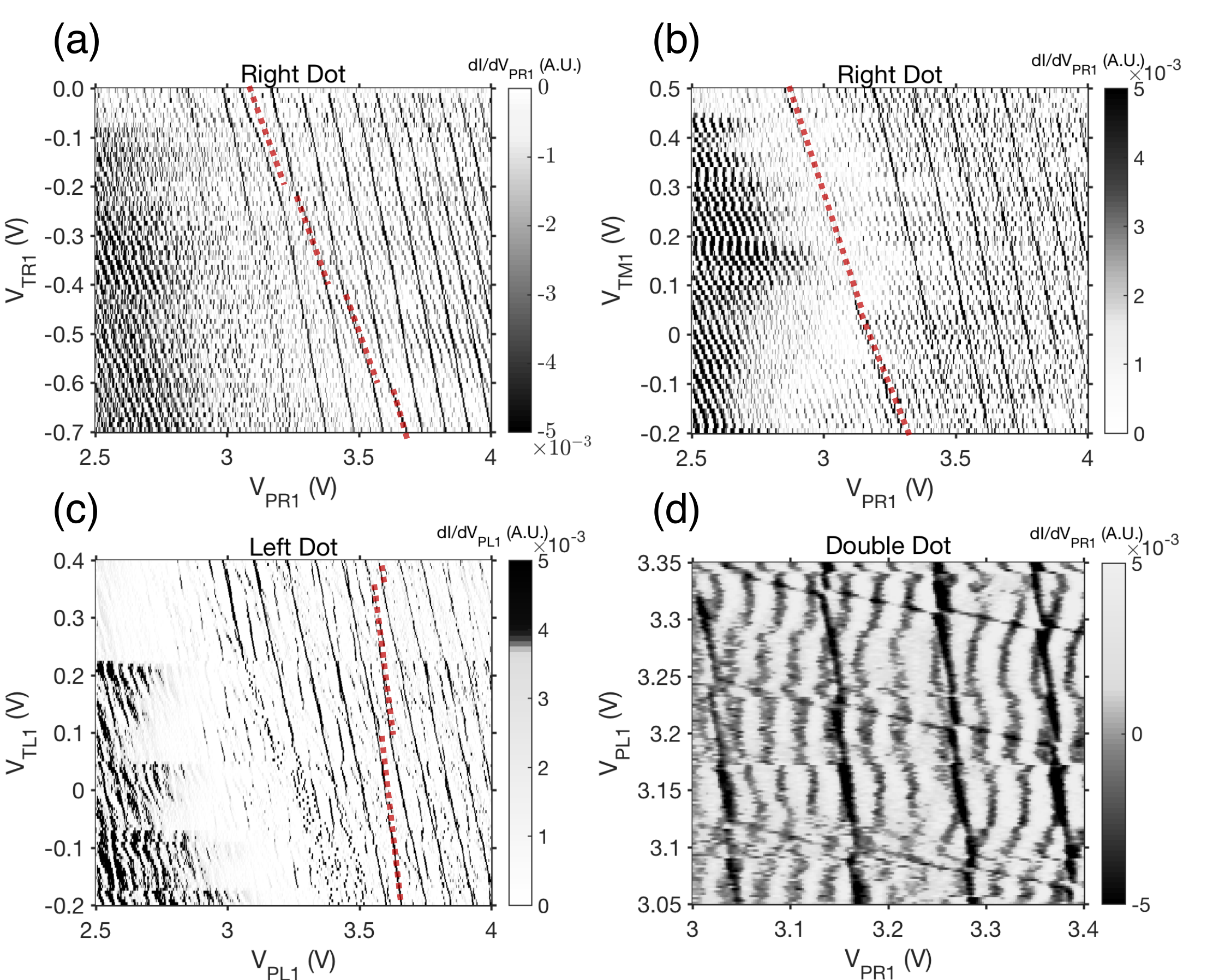}
\caption{\label{charge_sensing} Charge stability diagrams for the upper right dot (a), (b) and upper left dot (c), demonstrating the tuning of each dot down to 0 electrons. The regular parallel black lines indicate the quantum dot transitions. Three electron traps in the vicinity of the quantum dots are detected and are highlighted by the red dotted lines in (a)-(c). A persistent background oscillation is also present in the charge-sensing signal from the charging of individual poly-silicon grains in the poly-silicon depletion gate. (d) Demonstrates the controllable formation of a quantum double dot. The inter-dot coupling can be tuned via the tunnel gate TM$_1$ to merge the two dots into a single large dot.}
\end{figure}

Figure \ref{charge_sensing} shows the charge stability diagrams measured by the charge sensor dot of the upper two quantum dots as a function of the corresponding quantum dot plunger gate voltage and an adjacent tunnel barrier voltage. Three notable features are visible in Fig. \ref{charge_sensing} (a)-(c): a set of parallel electron transitions corresponding to the QD, three isolated individual lines corresponding to the charging of nearby defect states (highlighted in red), and a persistent background signal from the charging of poly-silicon grains.

The set of QD transitions are strongly coupled to the plunger gate voltages and their charging energies are consistent with the charging energy measured by bias spectroscopy. In addition, they show regularity over a large number of electron transitions, indicative of low interface disorder. As the tunnel barrier is made more negative (more opaque), the QD confinement increases and thus the spacing between QD transitions increases slightly. At sufficiently negative tunnel gate voltages, the tunnel barriers increase in opacity to the point where the tunnel rate can no longer keep up with the measurement scan rate and ``latching" behavior is seen in the lower region of the left-most transitions. Monitoring the transition point of the latching behavior for each electron transition ensures that the tunnel rate of each electron transition remains fast enough to be detected during the scan. In this way, electron transitions are not obscured by a tunnel rate being much slower than the measurement scan rate. Thus we demonstrate the depletion of each quantum dot to zero electrons.

Utilizing the charge sensor we may now extract the charging energies for the first electron transition, converting the plunger gate voltage difference between the first and second electron transitions to energy using the lever arms measured from the Coulomb blockade diamonds. We extract charging energies of 7.9 meV for the right dot and 7.6 meV for the left dot. Treating each dot as a metallic disc at the Si/SiO$_2$ interface, we can estimate the radius ($R$) of each dot in the single-electron regime from the dot's measured charging energy ($E_C$) and calculated self capacitance: $E_C=\frac{e^2}{C_{disc}}$, where $C_{disc}=8 \bar{\epsilon} \epsilon_{0}R$ \cite{zajac_dqd}. Here, $\bar{\epsilon}$ is given by $\frac{1}{2}(\epsilon_{Si}+\epsilon_{ox})$, the arithmetic mean of the relative permittivity of silicon ($\epsilon_{Si}$=11.7) and silicon dioxide ($\epsilon_{ox}$=3.9) \cite{ando}. This analysis yields radii of 37 nm for the right dot and 38 nm for the left dot, in good agreement with the effective lithographic radius of the dots (50x80 nm$^2$ $\rightarrow$ $R$=35.7 nm). 

In addition, bias-spectroscopy measurements of the left dot in the single-electron regime (not shown) reveal a conductance resonance 1.2 meV above the ground state which we attribute to the first excited orbital state. Treating the dot now as a 2D square box \cite{zajac_dqd}, we estimate the energy of the first excited state as $E_{orb}=\frac{3\hbar^2\pi^2}{2m^* L^2}$, where $m^*=0.19 m_0$ and $L$ is taken to be $(\pi R^2)^{1/2}$. This analysis yields $E_{orb}=1.3$ meV, in good agreement with the measured resonance. From these data, we conclude that these dots are lithographically defined and not dominated by random disorder at the Si/SiO$_2$ interface.

In the background of the charge sensing signal is a persistent signal caused by the charging of individual grains in the poly-silicon depletion gate. We determine these charging events to originate from the poly-silicon layer instead of the Si/SiO$_2$ interface disorder by three pieces of evidence. First, the transitions persist for all gate voltages probed and never deplete, indicating the charging object is a metallic island instead of a semiconductor dot. Second, the charging events are only evident in the charge sensing signal and not in the bias spectroscopy signal, indicating the object being charged does not originate from the Si/SiO$_2$ interface. The third and strongest piece of evidence is that the slope of these charging events turns positive when a charge sensing measurement is performed scanning the voltage of the poly-silicon gate against any other gate. A positive slope of these transitions is only possible when the object being charged is present on one of the gates being energized during the measurement. In this situation, the poly-silicon gate acts as both the electron reservoir and plunger gate to a poly-silicon grain acting as a quantum dot. Thus the Fermi level of the poly-silicon grain is pinned to the Fermi level of the surrounding poly-silicon gate. We emphasize that the poly-silicon transitions do not interact with or otherwise affect the charge transitions of the quantum dots, but simply add a noisy background signal. These ``phantom dot" signatures have been observed in other similar devices and can be eliminated by keeping the poly-silicon layer thicker than the order of the poly-silicon grain size.

The charge sensor in conjunction with the quantum dot can also be used as a local probe of defect states in the vicinity of the quantum dot. Defect states are evidenced by a single transition line slanted away from the set of parallel quantum dot transitions, highlighted in Fig. \ref{charge_sensing} (a)-(c), creating avoided crossing with the QD transitions. Scanning the QD gate voltages over a wide parameter range (limited to 4 V to avoid leakage between gates) to search for defects, we identify three distinct defect states, two in the vicinity of the right dot (highlighted in Fig. \ref{charge_sensing} (a) and (b)) and one in the vicinity of the left dot (highlighted in Fig. \ref{charge_sensing} (c)). These defect states are observed to hold only a single electron and the location of the defect state transition relative to the QD transitions are consistent from cool down to cool down. These observations suggest that the origin of these defects is a fixed positive charge in the oxide near the interface, as opposed to a mobile ionic charge which can freely migrate through the oxide at room temperature \cite{nicbrews}. Using the lithographic size of the dots, a rough estimate for the defect density can be obtained yielding $\approx3\times 10^{10}$ cm$^{-2}$. The estimate of the defect density from this method is consistent with the order of magnitude of the defect density measured by ensemble electron spin resonance measurements of the shallow trap density and conductivity measurements of the critical density of this material \cite{kim_sio2}. 

Finally, we demonstrate the controllable formation of a quantum double dot (Fig. \ref{charge_sensing} (d)). We observe the classic ``honey-comb" structure and triple-points. In this regime, there are several electrons in each quantum dot. The double-dot structure shown here demonstrates relatively weak inter-dot coupling, but we note that by tuning the tunnel barrier between the dots (TM$_1$), we are able to tune the inter-dot coupling from a single large dot to a double dot. We note that this double quantum dot enables a promising platform for future two-qubit and singlet-triplet qubit operations.

Another important device characteristic is charge noise. By biasing one of the dots to the edge of a Coulomb blockade peak, the dot is maximally sensitive to charge noise caused by fluctuations in the local electric field environment. In this measurement, a 1 mV DC bias is applied to the source of the device and the drain current is measured by the current preamplifier. The noise spectrum is then obtained from the output of the current preamplifier by an HP3561a signal analyzer. By comparing noise spectrum data at the edge of the Coulomb blockade peak with the noise spectrum taken at the top of the Coulomb blockade peak (where the device is minimally sensitive to local noise), the noise generated by the local environment can be extracted from noise generated by all other noise sources in the device and measurement circuitry \cite{jiang_noise, jung_noise}.  

\begin{figure}
\includegraphics[width=0.7\linewidth]{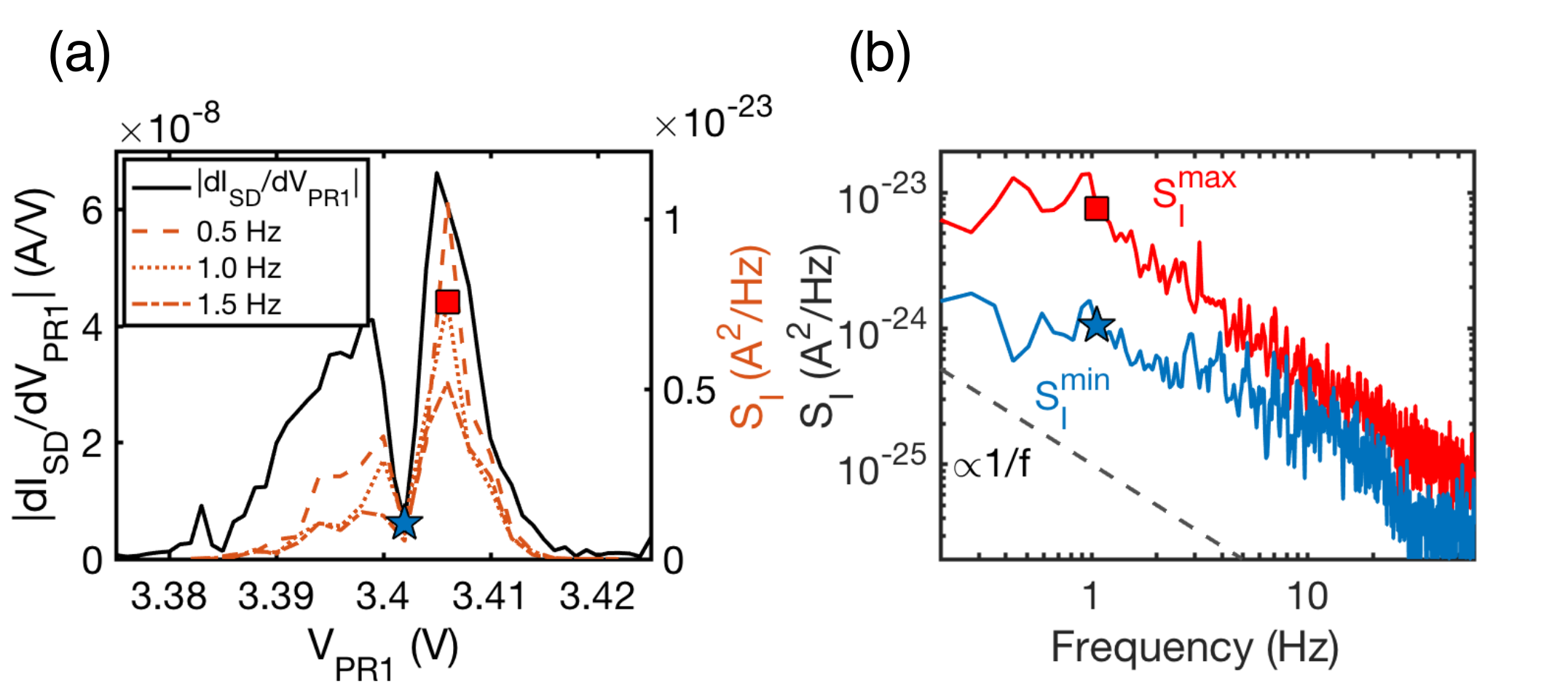}
\caption{\label{charge_noise} (a) Derivative of the QD drain current across a Coulomb blockade peak (left axis) overlaid with the power spectral density for 0.5, 1.0, 1.5 Hz (right axis). (b) Measured power spectral density at the maximally sensitive and minimally sensitive points of the QD Coulomb blockade peak (star and box indicated in (a)) as a function of frequency. 1/$f$ dashed line is shown for comparison.}
\end{figure}

Figure \ref{charge_noise}(a) shows the derivative of the quantum dot drain current with respect to the plunger voltage ($dI_{SD}/dV_{PR1}$) across a Coulomb blockade peak. Overlaid on the same plot is the magnitude of the noise spectrum at 0.5, 1.0, and 1.5 Hz. As expected, the magnitude of the noise correlates with the absolute value of the derivative of the device current, indicating the noise spectra are dominated by the fluctuations in the dot's chemical potential, $\epsilon$, from local environmental noise sources \cite{jiang_noise}. The noise spectra at the maximum and minimum points of $dI_{SD}/dV_{PR1}$, (indicated in Fig. \ref{charge_noise}(a) by $V_{S_I^{max}}$ and $V_{S_I^{min}}$) are plotted in Fig. \ref{charge_noise}(b) from 0.2 Hz to 59 Hz. The noise spectra follow a low frequency 1/$f$ dependence, frequently observed in electronic devices and indicative of an ensemble of two-level fluctuators in the vicinity of the device \cite{kirton_noise}. The measured magnitude of the noise spectra is well above the measured white-noise floor of our current preamplifier. 

We convert the measured current noise to the equivalent potential noise felt by the quantum dot, $\Delta\epsilon$, using the relation $\Delta I_\epsilon \alpha = |dI_{sd}/dV_{PR1}| \Delta\epsilon$ \cite{jiang_noise,jung_noise}. Here, $\Delta I_\epsilon = \sqrt{{S_I^{max}}-{S_I^{min}}}$. For $\Delta\epsilon$ evaluated at 1 Hz at 300 mK, we calculate a noise value of 3.4 $\micro$eV/Hz$^{1/2}$, consistent with other reported noise values reported at 1 Hz at 300 mK in MOS and Si/SiGe quantum devices \cite{jiang_noise}. This noise figure is likely to decrease at dilution refrigerator temperatures as has been observed in other work \cite{jiang_noise,dial2013}.

Finally, we perform magneto-spectroscopy measurements of the N=0$\rightarrow$1 and N=1$\rightarrow$2 electron transitions to measure the valley splitting. In bulk silicon, electrons reside in one of six degenerate valley states corresponding to the six symmetric minimums of the conduction band edge in $k$-space. When electrons are confined to an interface in a quantum dot, this six-fold degeneracy is lifted such that electrons preferentially populate the two valleys perpendicular to the interface \cite{das_sarma_valley}. The degeneracy of these remaining two valleys is referred to as the valley splitting and can be lifted through electric fields, confinement, and other details of the interface \cite{das_sarma_valley, gamble_vs}. For high fidelity spin selective quantum operations, the valley splitting must be large in relation to k$_B$T and the qubit energy \cite{das_sarma_valley}.

\begin{figure}
\includegraphics[width=0.8\linewidth]{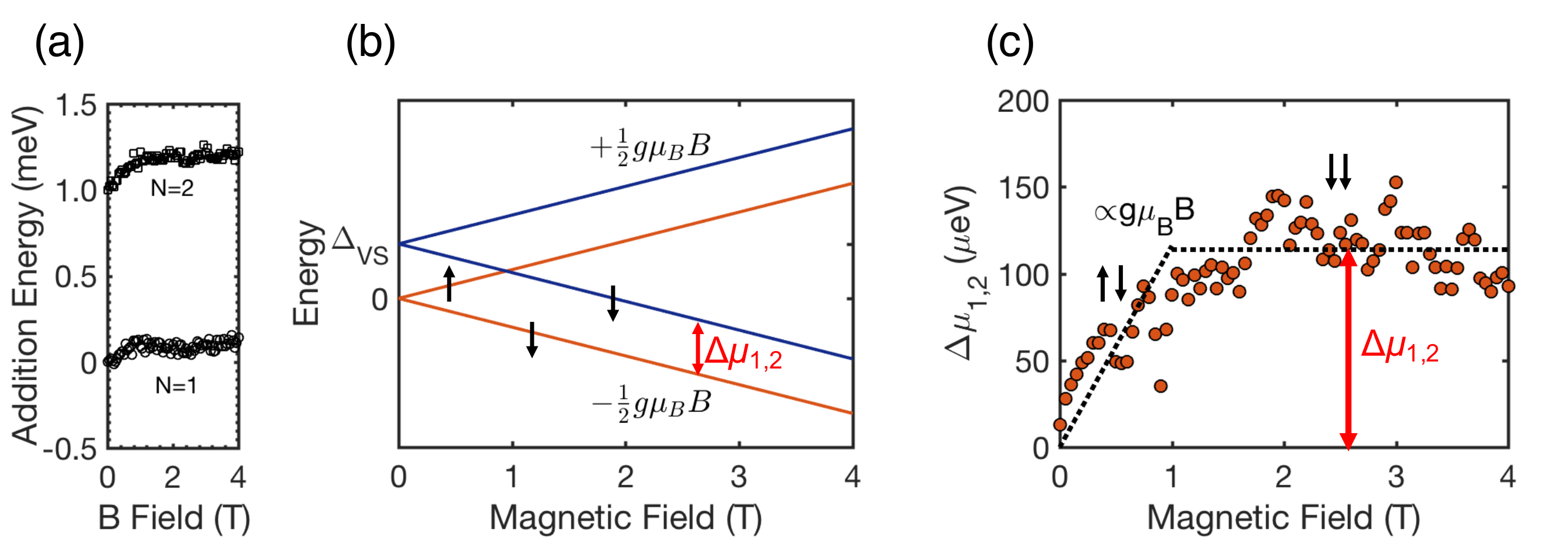}
\caption{\label{valley_splitting} (a) Evolution of of the conductance peaks of the first two electron transitions, offset by 1 meV for clarity, as a function of magnetic field. The evolution of these peaks is a combination of the orbital dependence on magnetic field, charge offset drift (very low frequency charge noise), and spin energies. (c) Cartoon of the spin energy levels of the two lowest valleys in silicon as a function of magnetic field, neglecting the effects of the magnetic field on the orbital energy. The difference in spin energies of the first and second electron is indicated by the red double-sided arrow. (d) Difference in addition energies of the first and second electron transitions ($\Delta\mu_{1,2}$). Near B=0, the data is fit to a slope of $g \mu_B$. A valley splitting of 110$\pm26$ $\micro$eV is measured.}
\end{figure}

Figure \ref{valley_splitting}(b) illustrates the energy diagram of the first two electrons loading into the quantum dot as a function of the magnetic field. The first electron loads into the lower valley state in the spin down configuration, and its addition energy decreases as a function of the magnetic field with a slope of $-\frac{1}{2}g\mu_B$=$-58$ $\micro$eV/T for $g$=2 (electron in Si), where the $\mu_B$ is the Bohr magneton. The second electron also loads into the dot in the lower valley state, but in the spin up configuration at low magnetic field, forming a spin singlet ground state, and its addition energy increases with a slope of $+\frac{1}{2} g \mu_B$. At a certain magnetic field value, the spin up arm of the lower valley state crosses the spin down arm of the upper valley state. Above this magnetic field, the second electron loads into a spin down configuration of the upper valley state, forming a spin triplet with the first electron. By subtracting the addition energy of the first electron from the second electron, we cancel for any changes in the dot's orbital energies \cite{borselli_vs} (which can be substantial in a perpendicular field) as well as slow charge offset drift (i.e. very low frequency drifts in the electrostatic environment on the order of 0.01$e$) \cite{hu2018,rudolph20018,hourdakis2008}. The difference in the transition energies of the first and second electron should first increase with a slope of $g\mu_B$ in the spin singlet configuration and then should be flat in the spin triplet configuration. 

In a perpendicular magnetic field, we tune the upper left dot to the first and second electron transitions, then monitor the position of each transition's Coulomb blockade peak as a function of magnetic field by fitting the conductance peak to a cosh$^{-2}$ function \cite{borselli_vs}. We convert the peak position from the plunger voltage to energy using the measured lever arm. In this experiment, we measure the conductance of the quantum dot instead of utilizing our charge sensor in order to bypass the noisy poly-silicon charging events which are apparent in the charge sensing signal. Figure \ref{valley_splitting}(a) plots the evolution of the conductance peaks of the first two electron transitions as a function of magnetic field. The evolution of these peaks contains contributions from the orbital and spin energies (which are both magnetic field dependent) as well as charge offset drift (which is time dependent, but independent of magnetic field and the same for both electron transitions). By taking the difference of the conductance peak positions, we may subtract out the orbital and charge offset drift components, leaving the difference in spin energies of the two electrons which is used to determine the valley splitting.

Figure \ref{valley_splitting}(c) shows the difference in addition energies of the first and second electrons. The energy difference increases initially with a slope $g \mu_B$, then saturates at a value of 110$\pm 26$ $\micro$eV, corresponding to the electron transition to a spin triplet in the upper valley state. Thus, we demonstrate a valley splitting large enough in this device to support spin-selective operations at typical dilution refrigerator temperatures ($\sim$100 mK = 8.7 $\micro$eV) \cite{das_sarma_valley}.

\section{Discussion}

In conclusion, we have engineered a low-disorder MOS quantum dot device and demonstrated a promising platform for electron spin qubits. We demonstrate the controllable formation of lithographically defined individual and double quantum dots with charging energies consistent with the lithographic size of the the dots. The local defect density around the quantum dots is low enough to support single electron occupation and correlates with ensemble measurements of the defect density as measured by ESR and percolation thresholds. Charge noise spectroscopy measurements show a 1/$f$ power spectral density yielding a value of 3.4 $\micro$eV/Hz$^{1/2}$, comparable to other Si MOS and Si/SiGe devices measured at 300 mK, and demonstrating a quiet noise environment. Finally, we measure a valley splitting of 110$\pm 26$ $\micro$eV, large enough to support high-fidelity spin operations. This work represents a platform for optimizing quantum dot disorder in MOS and a promising architecture for spin qubits.

%

\section{Availability of data, materials and methods}

The data presented in this work are available from the corresponding author upon reasonable request.

\section{Acknowledgements}

We gratefully acknowledge D. M. Zajac and J. R. Petta at Princeton University and R. M. Jock, M. Lilly and M. S. Carroll at Sandia National Laboratories for many useful discussions on device fabrication, instrumentation, and interpretation of data.

\section{Competing Interests}

The authors declare no competing interests.

\section{Funding}

This work was supported by the NSF through the MRSEC Program (Grant No. DMR-1420541). J.-S.K. is supported in part by the Program in Plasma Science and Technology at Princeton University. 

This work was performed, in part, at the Center for Integrated Nanotechnologies, an Office of Science User Facility operated for the U.S. Department of Energy (DOE) Office of Science by Los Alamos National Laboratory (Contract DE-AC52-06NA25396) and Sandia National Laboratories (Contract DE-NA-0003525).


%
%

%


\section{References}
\bibliography{references}

\end{document}